\documentclass[12pt]{amsart}
\usepackage[margin=0.8in]{geometry}

\usepackage{amsfonts}
\usepackage{amsthm}
\usepackage{amssymb}
\usepackage{times} 
\usepackage{graphicx}
\date{}

\newtheorem{theorem}{Theorem}

\newtheorem{corollary}[theorem]{Corollary}

\newtheorem{defn}{Definition}

\newtheorem*{example}{Example}

\begin{document}

\title{A short proof that all linear codes are weakly algebraic-geometric using Bertini theorems of B. Poonen}
\author{Srimathy Srinivasan}
\address{Department of Mathematics\\
University of Colorado\\
Boulder, CO 80309\\
USA}

\begin{abstract}
In this paper we give a simpler proof of  a deep theorem proved by Pellikan, Shen and van Wee that all linear codes are weakly  algebraic-geometric using a theorem of B.Poonen.

\end{abstract}
\maketitle

\section{Introduction}
\label{sec:curves}

Algebraic-geometric codes were first dicovered by Goppa (\cite{goppa}) and were further developed by Tsfasman,Vladut (\cite{agcodes}) and many others along the way. We start by briefly recalling the construction of algebraic-geometric codes.  The text \cite{hartshorne} by Hartshorne is a good reference for all the basic algebraic geometry  and notations we use in this  paper.

\begin{defn}
\label{defn:agcodes}
Let  $X$ be a smooth projective variety defined over $\mathbb{F}_q$ and let $\mathcal{P} = \{P_1, P_2, \cdots P_n\} \subseteq X(\mathbb{F}_q)$. Let $D$ be a divisor on $X$ such that the support of $D$ is disjoint from $\mathcal{P}$. Define  

\begin{align*}
L(D) = \{f \in \mathbb{F}_q(X)^*  | (f) + D \geq 0\} \cup \{0\}
\end{align*}
and consider the evaluation map:
\begin{align*}
Ev_\mathcal{P} : L(D)&\longrightarrow \mathbb{F}_q^n \\
 f&\longmapsto [f(P_1), f(P_2), ..., f(P_n)]
\end{align*}
The image of the map gives a linear code $\mathcal{C} = C_L(X, \mathcal{P},D)$ and we say that $\mathcal{C}$ is an algebraic-geometric code realized over $X$. 
\end{defn}

Given the data $X$, $\mathcal{P}$, $D$ as above,  let $\mathcal{L}$ denote the line bundle associated to $D$ and  $H^0(X,\mathcal{L})$ denote its global sections. Then we can get a code $C(X, \mathcal{P},\mathcal{L})$ equivalent to $C_L(X, \mathcal{P},D)$ as follows. First note that the local ring $\mathcal{L}_{P_i}$ modulo the maximal ideal of sections vanishing at $P_i$ denoted by $\overline{\mathcal{L}}_{P_i}$ is isomorphic to $\mathbb{F}_q$ by a choice of local trivialization. Then the image of the germ map  
\begin{align*}
\alpha_{\mathcal{P}}: H^0(X,\mathcal{L})&\longrightarrow \bigoplus_i^n \overline{\mathcal{L}}_{P_i}\cong \mathbb{F}_q^n 
\end{align*}
gives a linear code $C(X, \mathcal{P},\mathcal{L})$ that is same as the code $C_L(X, \mathcal{P},D)$   upto monomial equivalence. \\

\noindent \emph{Remark}: In Definition \ref{defn:agcodes}, if $X$ is a smooth curve, we get the original construction of Goppa (Goppa codes \cite{goppa}). In this case, the parameters of the code are easily estimated using the Riemann-Roch theorem. However, it is not so easy  for codes over higher dimensional varieties as  invoking Riemann-Roch  brings higher cohomology groups come into picture.\\
\indent In the paper \cite{pellikan}, Pellikan, Shen and van Wee define the notion of \emph{weakly algebraic-geometric codes} which we now recall:

\begin{defn}
\label{def:wag}
A  $q$-ary linear code $\mathcal{C}$  is said to be weakly algebraic geometric if there exists a projective non-singular absolutely irreducible curve $X$ defined over $\mathbb{F}_q$, $n$ distinct points $\mathcal{P} = \{P_1, P_2 \cdots P_n\}$ on $X$ and a divisor $D$ with support disjoint from $\mathcal{P}$ such that $\mathcal{C}= C_L(X, \mathcal{P}, D)$. 
\end{defn}

In their paper, the authors show that every linear code is weakly algebraic-geometric (Theorem 2, \cite{pellikan}). The goal of this paper is to give a simpler proof of this deep theorem using a theorem of B.Poonen. Although there are articles in the literature  such as \cite{alan} that apply Poonen's theorem to algebraic-geometric codes, there does not seem to be any literature that state this result.

\section{All linear codes are weakly algebraic-geometric}
In this section we give a shorter proof of  Theorem 2 of Pellikan, Shen, van Wee (\cite{pellikan}). \\

\indent We first show that algebraic-geometric codes are ubiquitous in the sense that every linear code can be realized  over some smooth variety. In fact we have the following stronger result.
\begin{theorem}
\label{thm:agcode}
 Let $\mathcal{C}$ be a linear code. Then  $\mathcal{C} = C_L(X,\mathcal{P},D)$ where $X$ is the blow up of some projective space at finitely many points,
$\mathcal{P}$ is a finite set of distinct $\mathbb{F}_q$-points in $X$ and $D$ is a divisor such that the support of $D$ is disjoint from $\mathcal{P}$.
\end{theorem}
\begin{proof}

 Let $\mathcal{C}$ be a $(n, k, d)_q$ linear code with $k \times n$ generator matrix $G$. Then the columns $C_1, C_2 , \cdots C_n$ of $G$ form (not necessarily distinct) points of $\mathbb{A}^k$. Then we can find an integer $r \geq 2$ and $n$ distinct points $P_1, P_2, \cdots P_n$  in $\mathbb{A}^{r+k}$ such that the projection map
\begin{align*}
\phi: \mathbb{A}^{r+k} &\rightarrow \mathbb{A}^k \\
[y_1, y_2, \cdots y_r, x_1, x_2 \cdots, x_k] &\rightarrow [x_1, x_2, \cdots, x_n]
\end{align*}
takes $P_i$ to $C_i$.   Let $y_0, y_1, \cdots, y_r, x_1, x_2, \cdots, x_k$ denote the coordinates of $\mathbb{P}^{r+k}$ .   Identify  $\mathbb{A}^{r+k}$  with the open affine set $y_0 = 1$ in $\mathbb{P}^{r+k}$.  For $1 \leq i \leq r$, let $V_i$ denote the point in $\mathbb{P}^{r+k}$ with $y_0 = y_i =1$ and all other coordinates $0$.  By choosing $r$ large enough we can assume that $V_i \neq P_j ~\forall i,j$. Let $X$ be the smooth geometrically integral variety obtained via the blow up  $\pi: X \rightarrow \mathbb{P}^{r+k}$ at the points  $V_i$  with the corresponding exceptional divisor $E_i$. Denote by $H$ the hyperplane section $y_0 = 0$ in $\mathbb{P}^{r+k}$. Then the global sections of the line bundle associated to the divisor $D= \mathcal{L}( \pi^* H - \sum_i  E_i)$ is generated by $x_0, x_1, \cdots, x_k$. It is  easy to see that  code $\mathcal{C} = C(X, \mathcal{P}, D)$ where $\mathcal{P}$ is the set $\{\pi^{-1} P_1,  \pi^{-1} P_2, \cdots, \pi^{-1} P_n\}$. 
\end{proof}

Let us  now restate the results on Bertini theorems over finite fields due to  B.Poonen. We refer the reader to Theorem 1.1 in \cite{poonen2} and remarks below Theorem 3.3 in \cite{poonen} for more details.

\begin{theorem}[Poonen]
\label{thm:poonen}
Let $X$ be a smooth, projective geometrically integral
  variety of $\mathbb{P}^n$ of dimension $m \geq 2$ over
  $\mathbb{F}_q$, and let $\mathcal{P} \subset X$ be a finite set of
  closed points. Then, given any integer $d_0$, there exists a hypersurface $H \subset
  \mathbb{P}^n$ of degree $d \geq d_0$ such that $ Y=H \cap X$ is
  smooth, projective and geometrically integral of dimension $m - 1$
  and contains $\mathcal{P}$.
\end{theorem}

\begin{example}
 Consider $X = \mathbb{P}^2$ over $\mathbb{F}_2$. Let $\mathcal{P}$ be the set of all 7 $\mathbb{F}_2$-points. Then, the curve $Y =  yz^3 + y^3 z + xy^3 + x^2 z^2 + x^2 y^2 + x^3 z$ is a smooth curve passing through $\mathcal{P}$. In fact, one can show that there are 24 smooth curves of degree 4 passing through $\mathcal{P}$.
\end{example}

Using Theorem \ref{thm:poonen} we now show that codes realized over higher dimensional varieties can be realized over curves.

\begin{theorem}
\label{thm:main}
Let $\mathcal{C}=C(X, \mathcal{P},\mathcal{L})$ be a code on a geometrically integral  smooth projective variety $X \subseteq \mathbb{P}^k$ of dimension $m \geq 2$ over $\mathbb{F}_q$. Then $\mathcal{C}$ can be realized over a smooth projective geometrically integral curve. In particular, there exists a geometrically integral smooth projective curve $Z$ containing $\mathcal{P}$  such that $\mathcal{C} =  C(Z, \mathcal{P}, \mathcal{L}|_Z)$.
\end{theorem}

\begin{proof}
 Let $ Y=H \cap X$ be as Theorem \ref{thm:poonen} with degree $d$ of $H$ large enough and let $i: Y \hookrightarrow X$ denote the inclusion morphism.  Then, we have the following short exact sequence on $X$.
\begin{align*}
 0 \longrightarrow \mathcal{I}_Y \longrightarrow \mathcal{O}_X \longrightarrow i_*\mathcal{O}_Y \longrightarrow 0
\end{align*}
where  $\mathcal{I}_Y = \mathcal{O}_X(-d)$. Tensoring with $\mathcal{L}$ we get
\begin{align*}
0 \longrightarrow  \mathcal{L} \otimes \mathcal{I}_Y \longrightarrow \mathcal{L} \longrightarrow \mathcal{L}\otimes i_*\mathcal{O}_Y \longrightarrow 0
\end{align*}
(Here tensoring is over $\mathcal{O}_X$). The above short exact sequence  gives rise to a long exact sequence in cohomology on $X$
\begin{align*}
 0 \longrightarrow  H^0(\mathcal{L} \otimes \mathcal{I}_Y, X) \longrightarrow H^0(\mathcal{L}) \longrightarrow H^0(\mathcal{L}\otimes i_*\mathcal{O}_Y, X) \longrightarrow  H^1(\mathcal{L} \otimes \mathcal{I}_Y, X) \longrightarrow \cdots
\end{align*}
By duality, we have 
\begin{align*}
H^i(\mathcal{L} \otimes \mathcal{I}_Y, X) \simeq H^{m-i} (\omega_X \otimes \mathcal{I}_Y^\vee \otimes \mathcal{L}^\vee, X)  \simeq H^{m-i} (\omega_X \otimes  \mathcal{L}^\vee \otimes \mathcal{O}_X(d), X) 
\end{align*}
where $\omega_X$ is the canonical sheaf on $X$. Since $\mathcal{O}_X(1)$ is ample, for large enough $d$, $H^0(\mathcal{L} \otimes \mathcal{I}_Y, X)$ and $H^1(\mathcal{L} \otimes \mathcal{I}_Y, X)$ vanishes and we get a canonical isomorphism obtained via restriction
\begin{align*}
 H^0(\mathcal{L}) &\tilde{\longrightarrow} H^0(\mathcal{L}\otimes i_*\mathcal{O}_Y, X) \simeq H^0(\mathcal{L}|_{Y} \otimes \mathcal{O}_Y, Y) \simeq H^0(\mathcal{L}|_Y, Y)\\
f & \longmapsto  f_{|Y}.
\end{align*}
Inducting the above argument by replacing the $m$-dimensional variety $X$ with $(m-1)$-dimensional variety $Y$ and $\mathcal{L}$ with $\mathcal{L}|_Y$ we get the result.
\end{proof}

Hence, given a code over a geometrically integral smooth projective variety, we have realized it over a  geometrically integral smooth projective curve.  As a consequence we get:
\begin{corollary}[Pellikan, Shen, van Wee]
\label{cor:wag}
All linear codes are weakly algebraic-geometric.
\end{corollary}
\begin{proof}
This easily follows from Theorem \ref {thm:agcode} and Theorem \ref{thm:main}.
\end{proof}

\renewcommand{\abstractname}{Acknowledgements}
\begin{abstract}
I would like to thank Patrick Brosnan and  Lawrence Washington for their valuable comments and discussions. I would also like to thank Richard Rast for writing a computer program to verify some results.
\end{abstract}

\bibliographystyle{alpha}
\bibliography{short_proof_poonen}

\def\cprime{$'$}
\begin{thebibliography}{PSvW91}

\bibitem[Cou11]{alan}
Alain Couvreur.
\newblock Differential approach for the study of duals of algebraic-geometric
  codes on surfaces.
\newblock {\em J. Th\'eor. Nombres Bordeaux}, 23(1):95--120, 2011.

\bibitem[Gop81]{goppa}
V.~D. Goppa.
\newblock Codes on algebraic curves.
\newblock {\em Dokl. Akad. Nauk SSSR}, 259(6):1289--1290, 1981.

\bibitem[Har77]{hartshorne}
Robin Hartshorne.
\newblock {\em Algebraic geometry}.
\newblock Springer-Verlag, New York-Heidelberg, 1977.
\newblock Graduate Texts in Mathematics, No. 52.

\bibitem[Poo04]{poonen}
Bjorn Poonen.
\newblock Bertini theorems over finite fields.
\newblock {\em Ann. of Math. (2)}, 160(3):1099--1127, 2004.

\bibitem[Poo08]{poonen2}
Bjorn Poonen.
\newblock Smooth hypersurface sections containing a given subscheme over a
  finite field.
\newblock {\em Math. Res. Lett.}, 15(2):265--271, 2008.

\bibitem[PSvW91]{pellikan}
R.~Pellikaan, B.-Z. Shen, and G.~J.~M. van Wee.
\newblock Which linear codes are algebraic-geometric?
\newblock {\em IEEE Trans. Inform. Theory}, 37(3, part 1):583--602, 1991.

\bibitem[TV91]{agcodes}
M.~A. Tsfasman and S.~G. Vl{\u{a}}du{\c{t}}.
\newblock {\em Algebraic-geometric codes}, volume~58 of {\em Mathematics and
  its Applications (Soviet Series)}.
\newblock Kluwer Academic Publishers Group, Dordrecht, 1991.
\newblock Translated from the Russian by the authors.

\end{thebibliography}

\end{document}